\documentclass[sigconf,nonacm]{acmart}
\usepackage{listings}
\AtBeginDocument{%
  }

\DeclareMathOperator*{\argmax}{argmax} 

\usepackage{subcaption}
\usepackage{xcolor}

\begin{document}


\title{Automated Query-Product Relevance Labeling using Large Language Models for E-commerce Search}




\author{Jayant Sachdev}
\authornotemark[2]
\email{jayant.sachdev@walmart.com}
\affiliation{%
  \institution{Walmart Global Tech}
  \city{Sunnyvale}
  \state{CA}
  \country{USA}
  \postcode{94086}
}

\author{Sean  D Rosario}
\authornote{Corresponding author}
\authornote{These authors contributed equally to this research.}
\email{sean.drosario@walmart.com}
\affiliation{%
  \institution{Walmart Global Tech}
  \city{Sunnyvale}
  \state{CA}
  \country{USA}
  \postcode{94086}
}

\author{Abhijeet Phatak}
\authornotemark[2]
\email{abhijeet.phatak@walmart.com}
\affiliation{%
  \institution{Walmart Global Tech}
  \city{Sunnyvale}
  \state{CA}
  \country{USA}
  \postcode{94086}
}
\author{He Wen}
\email{vicky.wen@walmart.com}
\affiliation{%
  \institution{Walmart Global Tech}
  \city{Sunnyvale}
  \state{CA}
  \country{USA}
  \postcode{94086}
}
\author{Swati Kirti}
\email{swati.kirti@walmart.com}
\affiliation{%
  \institution{Walmart Global Tech}
  \city{Sunnyvale}
  \state{CA}
  \country{USA}
  \postcode{94086}
}
\author{Chittaranjan Tripathy}
\email{ctripathy@walmart.com}
\affiliation{%
  \institution{Walmart Global Tech}
  \city{Sunnyvale}
  \state{CA}
  \country{USA}
  \postcode{94086}
}

\definecolor{LLM4}{HTML}{1f77b4}
\definecolor{LLM1}{HTML}{ff7f0e}
\definecolor{LLM3}{HTML}{2ca02c}
\definecolor{LLM5}{HTML}{d62728}
\definecolor{LLM2}{HTML}{9467bd}
\newcommand{\usellmfour}[1]{{\color{LLM4}{#1}}}
\newcommand{\usellmone}[1]{{\color{LLM1}{#1}}}
\newcommand{\usellmtwo}[1]{{\color{LLM2}{#1}}}
\newcommand{\usellmfive}[1]{{\color{LLM5}{#1}}}
\newcommand{\usellmthree}[1]{{\color{LLM3}{#1}}}
\begin{abstract}
{Accurate query-product relevance labeling is indispensable to generate ground truth dataset for search ranking in e-commerce. Traditional approaches for annotating query-product pairs rely on human-based labeling services, which is expensive, time-consuming and prone to errors. In this work, we explore the application of Large Language Models (LLMs) to automate query-product relevance labeling for large-scale e-commerce search. We use several publicly available and proprietary LLMs for this task, and conducted experiments on two open-source datasets and an in-house e-commerce search dataset. Using prompt engineering techniques such as Chain-of-Thought (CoT) prompting, In-context Learning (ICL), and Retrieval Augmented Generation (RAG) with Maximum Marginal Relevance (MMR), we show that LLM's performance has the potential to approach human-level accuracy on this task in a fraction of the time and cost required by human-labelers, thereby suggesting that our approach is more efficient than the conventional methods. We have generated query-product relevance labels using LLMs at scale, and are using them for evaluating improvements to our search algorithms. Our work demonstrates the potential of LLMs to improve query-product relevance thus enhancing e-commerce search user experience. More importantly, this scalable alternative to human-annotation has significant implications for information retrieval domains including search and recommendation systems, where relevance scoring is crucial for optimizing the ranking of products and content to improve customer engagement and other conversion metrics.}
\end{abstract}

\ccsdesc[300]{Computing methodologies~Information extraction}
\ccsdesc[500]{Information systems~Relevance assessment}
\ccsdesc[100]{Information systems~Search engine indexing}
\ccsdesc[500]{Information systems~Novelty in information retrieval}
\ccsdesc[500]{Computing methodologies~Natural language generation}
\ccsdesc[100]{Computing methodologies~Language resources}

\keywords{Large Language Models, Search Relevance, Information Retrieval, Data Annotation, Prompt Engineering, Retrieval Augmented Generation, Maximum Marginal Relevance.}

\maketitle
\newcommand{\usered}[1]{{\color{red}{#1}}}
\newcommand{\usergreen}[1]{{\color{green}{#1}}}
\newcommand{\useblue}[1]{{\color{blue}{#1}}}
\newcommand{\useorange}[1]{{\color{orange}{#1}}}

\section{Introduction}
When a customer performs a search on an e-commerce platform like Walmart, it is essential that the search results show the exact products or relevant products they are looking for, saving the customer time and effort while providing a positive shopping experience.
The product search experience on e-commerce platforms is powered by a collection of machine learning models and algorithms across the stages of pre-retrieval, retrieval, and post-retrieval ranking and filtering \cite{pre-training-tasks}. Many of these models rely on accurate query-product (Q-P) relevance label data for training and testing. Specifically, these labels  indicate how relevant a given product is to a customer search query. For example, the product ``\texttt{Ergonomic PU leather high back office chair with flip-up armrest}'' is highly relevant to the search query  ``\texttt{leather chair}'', while the product ``\texttt{hard floor beveled edge chair mat}'' is not relevant to the same query.
\begin{figure}[!h]
\centering
\minipage{0.81\columnwidth}%
    \includegraphics[height=0.5 \linewidth,width=1.1\linewidth]{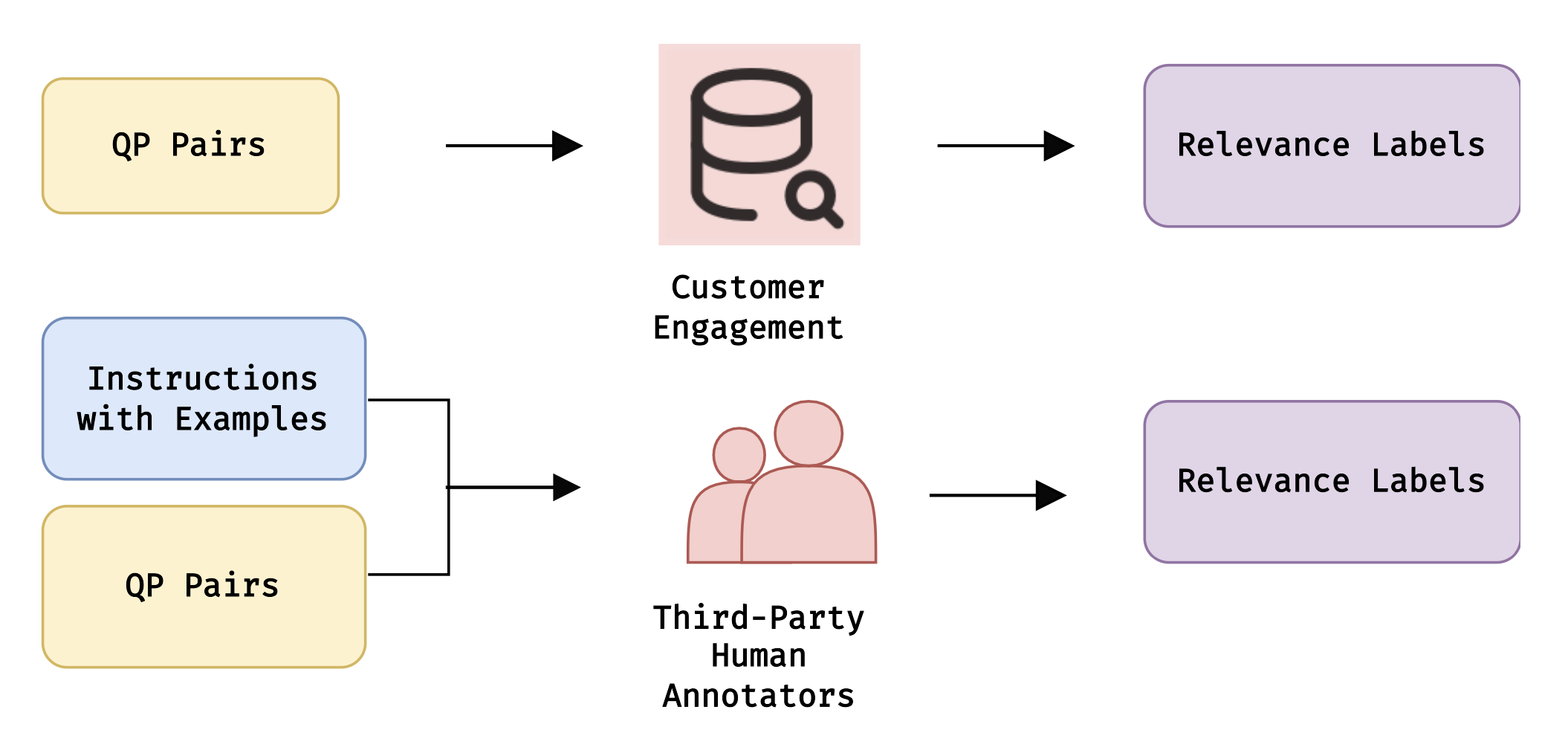}
    \caption*{(a) Previous approaches of acquiring Q-P relevance}\label{fig:prior_methods}
\endminipage\hspace{5mm}%
\minipage{0.9\columnwidth}%
    \includegraphics[width=0.95\linewidth]{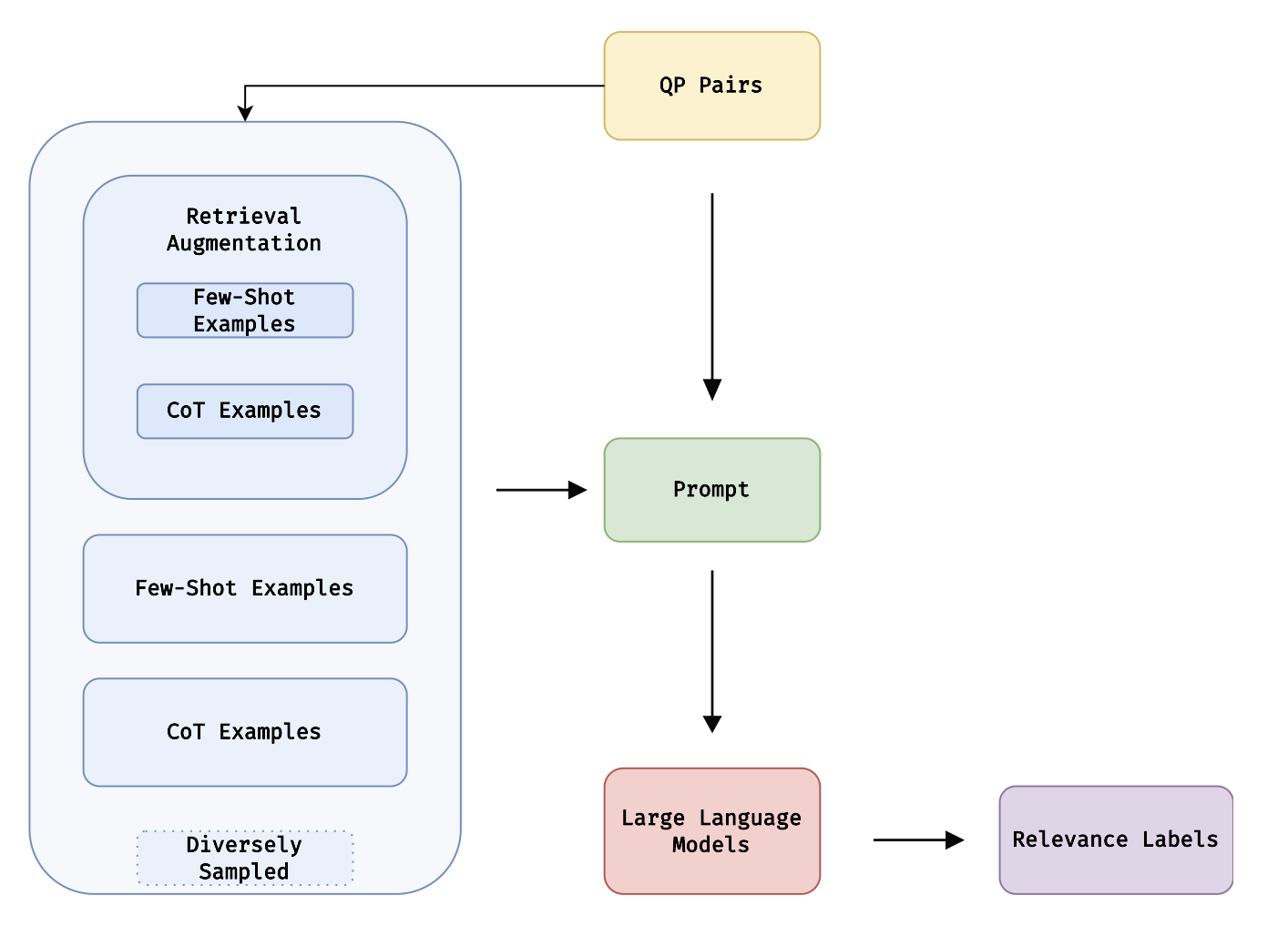}
    \caption*{(b) Our approach to annotating Q-P relevance}\label{fig:current_methods}
\endminipage \\
\caption{Comparison of pipeline architectures for Q-P relevance annotation}
\label{fig:comparison_of_methods}
\end{figure}

\begin{figure*}[!h]
    \centering
    \includegraphics[width=1.75\columnwidth]{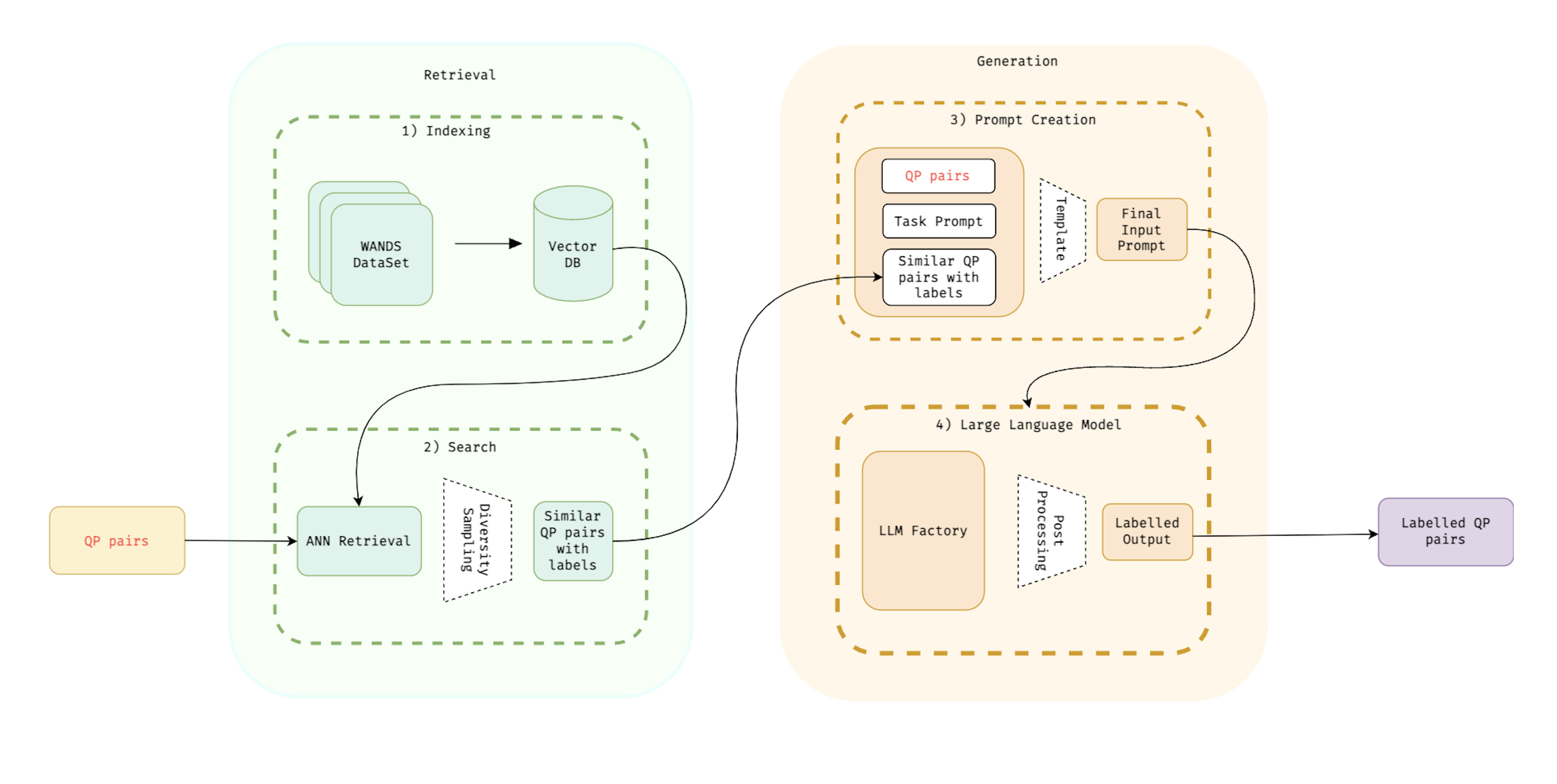}
    \caption{System Architecture for Automated Q-P relevance with LLM and RAG}
    \label{fig:RAG}
\end{figure*}

We have worked with an external data labeling service which employs human labelers. The labelers were given Q-P pairs with detailed instructions and examples to guide labelers. They were not given any information on customer engagement data for Q-P pairs. Typically, three to five independent assessments are gathered per Q-P pair. Despite the guidelines, human labelers do not always reach consensus on an annotation, hence labeled data is often noisy. Furthermore, external human labeling is expensive and time-consuming, and hence only Q-P pairs corresponding to a small fraction of all queries can be labeled. Despite these challenges, external human labelers are considered as the most reliable method of acquiring data annotations, and is widely used across e-commerce and internet industries.
Previously, Q-P relevance labels were obtained from clickstream data and human annotation as shown in Figure \ref{fig:comparison_of_methods}(a). Higher engagement of a product given a search query would be considered a positive signal for relevance. However, cold-start products (such as newly introduced products and tail products that are sold infrequently) may lack sufficient impressions, click-through-rate (CTR), or add-to-cart (ATC) rate information due to their limited exposure to the customers. Furthermore, some products might get a lot of impressions and clicks even if they are not very relevant to the search query. For example, the product ``\texttt{fresh lime}'' gets a relatively high number of impressions and clicks for the query ``\texttt{lemon}'', but it is not a very relevant product.
Identifying the strong need for an automated framework for Q-P relevance labeling, in this work, we demonstrate that Large Language Models (LLMs) can be used successfully to obtain accurate relevance labels in a fully automated manner and at scale. A schematic of our approach is shown in Figure \ref{fig:comparison_of_methods}(b), which we will detail in the subsequent sections. We have experimented with several LLMs for this task. We have explored  different e-commerce search datasets, as well as ablation studies on the effectiveness of prompt engineering techniques including Chain-of-Thought (CoT) prompting, In-context Learning (ICL), and Retrieval Augmented Generation (RAG) with Maximum Marginal Relevance (MMR). Figure \ref{fig:RAG} shows Retrieval-Augmented Generation (RAG) architecture, designed and used in our work, which contains a retrieval component that retrieves relevant labeled query-product pairs for a given query-product pair. These relevant query-product pairs, providing the contextual information, are then used to create prompts which are provided as input to the LLMs. 
\section{Background}
This work is an application of data-centric Artificial Intelligence (AI), which is systematic engineering of the data used to build an AI system, including but not limited to automated data labeling, collection, augmentation, and cleaning \cite{zha2023datacentric}. One approach to automated data labeling is called \textit{programmatic labeling} which includes using human-defined labeling functions (LFs) that capture labeling logic to auto-generate large training sets \cite{Ratner2017SnorkelRT}. Programmatic labeling significantly increases efficiency over manual labeling through automating the labeling process. However, labeling decisions still require human domain expertise on what heuristics should be applied \cite{guan2023large}.
\subsection{Large Language Models}
Large Language Models can generate human-like text and demonstrate state-of-the-art logical reasoning ability across a broad range of reasoning tasks and benchmarks \cite{Achiam2023GPT4TR}. With advancements in the field, open-source models have also proven to be an effective alternative to GPT models in multiple reasoning tasks \cite{vergho2024comparing, llama, mistral, mixtral}. Prior work has shown that LLMs can be used to label data at par with human labeling on a variety of NLP tasks, with up to 96\% saving in costs \cite{wang-etal-2021-want-reduce}. LLMs are proven to be a viable substitute to crowd-sourced annotators and can match or surpass crowd-sourced annotators \cite{he2023annollm}. Recent work has also explored the idea of using LLMs to annotate shopping relevance \cite{soviero2024chatgpt,faggioli2023perspectives}.
\subsection{Mathematical Formulation}
Let $(Q, P)$ denote the set of all possible Q-P pairs. We define the task of relevance labeling as a classification problem with the candidate classes set $\mathcal{Y} = \{y_1, y_2, \ldots, y_n\}$. We want to find a map $\Phi_h: (Q,P) \rightarrow \mathbb{R}^{\mathcal{Y}}$ that is able to ideally score the relevance for the Q-P pairs against the candidate classes. We assume $Q$ and $P$ are infinite sets \cite{bagheri-garakani-etal-2022-improving}. Given an input query $q \in Q$ and product $p \in P$, we want to maximize the probability $$P(y_{pred}=y_{true}|q,p),$$ where \[y_{pred} = \argmax_{y_{i} \in \mathcal{Y}} \Phi_h(q, p).\]
$\Phi_h$ can be expressed as a function of the choice of the LLM and associated sampling parameters ($\mathcal{L}$) along with the baseline input prompt or context ($\mathcal{P}$) to the LLM. 
With advanced prompting techniques, we modify the context to include additional instructions $I$ and set of $k$ few shot examples or demonstrations that are added to the input context provided to the LLM. This can be represented as $\mathcal{P}_{I,k}$, and we can express
$$\Phi_h = \Psi(\mathcal{L}, \mathcal{P}_{I,k}),$$
where $\Psi(\cdot, \cdot)$ is a function of $\mathcal{L}$ and $\mathcal{P}_{I,k}$.
In case of Chain-of-Thought (CoT) prompting, we include additional instructions  to step by step reason through the generation process that is $\mathcal{I}_{CoT}$. With crowd-sourced data annotation, we first need to provide the annotators with a thorough description of the labeling task, with meanings of each category and specific examples with explanations. Therefore, using the same approach, we guide LLMs to annotate data by providing the task description and some labeled examples. We observe that prompting the LLM to explain the reasoning behind the provided annotation can improve the labeling accuracy. We use RAG to retrieve most similar and hence most relevant demonstrations or examples that are represented as $k_{RAG}$. Since we use the retriever to provide additional context simply by concatenating the context with these examples. RAG exhibits significant potential in enhancing the quality of prompts by providing external knowledge as a preliminary step before generating the response, thereby ensuring contextual precision and a higher degree of detail. We use Maximum Marginal Relevance (MMR) \cite{mmr} retrieval to iteratively find documents that are dissimilar to previous results. MMR has been shown to improve retrievals in LLMs \cite{mmrinllm}. Consider the set $D$ consisting of all candidate documents and $R$ consisting of the previously chosen (retrieved) documents, and $q$ representing a query. We also consider two similarity functions -- $Sim_1$ that compares a document with the query, and $Sim_2$ that assesses the similarity between two documents. Let $d_i$ and $d_j$ denote individual documents in $D$ and $R$.  Then MMR is given by the following formula:
\begin{small}
$$\text{MMR}=\arg\max_{d_i \in D \setminus R} [ \lambda_{MMR} \cdot Sim_1(d_i, q) - (1 - \lambda_{MMR}) \cdot \max_{d_j \in R} Sim_2(d_i, d_j) ],$$
\end{small}
where the first term and the second term respectively represent relevance and diversity. The adjustable parameter $\lambda_{MMR}$ balances the importance of relevance and diversity. $\lambda_{MMR} \rightarrow 1$ prioritizes relevance whereas $\lambda_{MMR} \rightarrow 0$ prioritizes diversity.
Finally, we select the optimal map $\hat{\Phi}_{h}$ through experiments on ground truth data and ablation studies.
\section{Methods}
\subsection{Prompt Engineering}
Prompt engineering is the systematic design and optimization of input prompts to guide the LLM responses, ensuring accuracy, relevance, and coherence in the generated output \cite{chen2023unleashing}.
\subsubsection{In-Context Learning (Few-Shot Prompting)}
In-context learning refers to a setting in which the model is given a few demonstrations of the task at inference time, but no weight and gradient updates are allowed \cite{icl, dong2023survey}. The major advantage of this technique is reduced reliance on task-specific data and does not require fine-tuning on domain specific datasets.
\subsubsection{Chain-of-Thought (CoT) Prompting}
CoT prompting can enhance reasoning in language models by decomposing problems into intermediate steps \cite{wei2023chainofthought}. It provides an interpretable insight into the model's behavior, revealing how it might have reached a certain answer and helping identify where the reasoning may have gone wrong. Using CoT, we provided the reasoning as to why certain label was selected for each of the examples used in few-shot prompting. There are multiple ways to invoke a CoT response. In our experiments we use the words \texttt{'Let's think step by step'} in the few-shot examples to invoke CoT responses \cite{kojima2023large}.
\subsubsection{Retrieval Augmented Generation (RAG)}
RAG technique utilizes input to identify relevant information for the task at hand, which is then used to offer additional context in the formation of input prompt to the LLMs. The RAG architecture contains two main components: a retriever that identifies relevant information (documents) in response to input query, and a generator that generates the ouput, taking into account the context of retrieved information, the initial query, and required task information. It has been noted that extensive pre-trained language models have the capacity to encapsulate factual knowledge within their parameters and can attain superior outcomes when they are fine-tuned for distinct NLP tasks. Despite this, their proficiency in accessing and manipulating knowledge precisely is still lacking, leading to less than satisfactory performance in tasks that are heavily reliant on knowledge\cite{rag}. As shown in Figure \ref{fig:RAG}, given a query-product pair, using the RAG technique, we provide the relevant context information about the query-product pair to the LLMs, which proves to be beneficial in generating the correct relevance labels.  
\subsubsection{Maximum Marginal Relevance Retrieval (MMR)}
Retrieving similar Q-P pairs as few shot examples with RAG often leads to a set of examples with a high degree of overlap. Hence the amount of new information that each example adds to the prompt is limited. In order to overcome this, we added a component of diversity into the few-shot examples retrieved, as described in the previous section. MMR ensures that examples retrieved are both relevant to the question Q-P pair, while being sufficiently diverse. We experiment with different values of $\lambda_{MMR}$ to find the optimal balance between similarity and diversity. 
\subsection{Details of our Prompt}
Our prompt consists of three parts. First, we have a context prompt which specifies the instruction $I$ to the LLM including the classes relevance labels with explanations of each label and the expected output structure. Second, we do in-context learning by providing sample inputs (search query and product title) and the outputs (corresponding relevance labels). For CoT, we included more instructions in the instruction prompt $I$ and provided annotated few-shot examples. Lastly, we provide the Q-P actual pairs we want labeled by the LLM (see Figure \ref{fig:RAG} for a schematic).

\subsection{Datasets}
We perform experiments on two publicly available datasets and also on our proprietary dataset to validate the effectiveness of our methods.\\
\textbf{ESCI Dataset: }
ESCI dataset \cite{reddy2022shopping} is an e-commerce dataset that has Q-P pairs that are human-annotated with corresponding relevance labels. We assume that the annotations provided are accurate, and serve as ground truth labels. There are 4 classes for relevance labels: \texttt{Exact}, \texttt{Substitute}, \texttt{Complement}, and \texttt{Irrelevant}. We sampled 5000 Q-P pairs from the test split of the ESCI dataset, with an even split across the four class labels to create a test set for experimentation and evaluation purposes.\\
\textbf{WANDS Dataset: }
WANDS \cite{wands} is another e-commerce dataset that contains Q-P pairs that are human-annotated with the corresponding relevance labels. We assume that the annotations provided are accurate, and serve as ground truth labels. There are 3 classes for relevance labels: \texttt{Exact}, \texttt{Partial}, and \texttt{Irrelevant}. Since there is no train/test split in the data, we sampled 5000 Q-P pairs evenly across the three class labels to create a test set for experimentation and evaluation purposes.\\
\textbf{Walmart Mexico Search Dataset: }
Our primary data source is Walmart Mexico search session data, which contains customer search queries and product impressions. This data is proprietary and acquired from Walmart's internal databases. We sampled 100 queries uniformly across search traffic segments. For each of the 100 sampled queries, we sample 10 products retrieved before re-ranking. Thus, we have 1000 distinct query product pairs in our ground truth data, mostly in Spanish.
Domain experts have manually annotated relevance across the 1000 Q-P pairs. The ground truth data has 5 relevance classes: \texttt{Excellent}, \texttt{Good}, \texttt{Okay}, \texttt{Bad}, and \texttt{Embarrassing}. The data is heavily skewed towards \texttt{Excellent} because the products per query are sampled from search retrieval data, which is likely to be more relevant than not. Hence, for this dataset, weighted F1 score is a more reliable metrics than average F1 score.\\
\subsection{Experimental Setup}
Across all three datasets, we experimented with 8 Few-Shot examples and 16 Few-Shot examples for in-context learning. These examples could be randomly sampled from the training set (referred to as FS in the tables and figures), or retrieved based on embedding distance from the test question (referred to as FS\_RAG in the tables and figures), and diversity could be introduced to the RAG examples (referred to as FS\_RAG\_MMR in tables and figures). For experiments with Walmart Mexico data, we sampled some Q-P pairs outside the test set from historical search relevance data to be used as Few-Shot (FS) examples. Explanations for few shot example were also manually annotated for CoT examples. Third-party human annotations serve as the baseline for experiments on this data. For the ESCI dataset, FS examples were randomly sampled from the train split of the ESCI dataset.
For the prompt, we made use of descriptions of relevance labels as defined in the \cite{wands,reddy2022shopping} for the experiments on WANDS and ESCI respectively. We used an 80GB GPU for running inference of open source LLMs: LLM1 is an 8 billion parameter LLM \cite{llama}, and LLM2 is a 70 billion parameter version of the same model. LLM3 is another open-source model with 7 billion parameters \cite{mistral}, and LLM4 \cite{mixtral} is a Mixture-of-Experts model based on 8 models of LLM3.   We used \texttt{vLLM} \cite{vLLM} that uses PagedAttention to speed up the inference to make our approach more scalable and resource friendly. We used Activation-aware Weight Quantized (AWQ) models \cite{lin2023awq} that could not be fit in a single 80GB GPU like LLM2 and LLM4. LLM5 \cite{gemini} is not open-sourced, so we used a paid API provided to perform the LLM calls. To implement RAG as shown in Figure \ref{fig:RAG}, we used \texttt{ChromaDB} \cite{chromadb} as our vector store for retrieving similar query-product pairs from training set. We then use this retrieved query-product pairs along with our task prompt, and input query-product pair to create an input prompt for the LLM. We used \texttt{scikit-learn} \cite{sklearn} implementations for computing evaluation metrics that are described in the subsequent section in detail.
\subsection{Evaluation Metrics}
We compute the following metrics across LLM predictions on the aforementioned datasets.\\
\textbf{Accuracy:} $\texttt{Accuracy}(y, \hat{y}) = \frac{1}{n_\text{samples}} \sum_{i=1}^{n_\text{samples}} \mathbf{1}(\hat{y}_i = y_i)$ ,where $\hat{y}_i$ is the predicted value of the 
$i$-th sample and $y_i$ is the corresponding true value, then the fraction of correct predictions over all $n_{samples}$ gives us the accuracy score. $\mathbf{1}(x)$ is the indicator function.\\
\textbf{Average $F_{1}$ Score:} $F_{1} = 2 \* \frac{P\*R}{P+R}$, where $P$ is Precision and $R$ is Recall. This measures the prediction accuracy across all relevance classes without accounting for class imbalance. Higher Average $F_{1}$ score means better classification performance.\\
\textbf{Weighted $F_{1}$ Score: } $F_{1}^{\omega} = \sum_{j=1}^{n_i} \hat{w}_i F_1(i)$ To compute the weighted $F_{1}$ score the $F_{1}$ score for each class is calculated individually, and averaged using normalized support weights per label. This can result in a better measure than average $F_{1}$ score, as it accounts for class imbalance. Higher weighted $F_{1}$ score means better classification performance.
\section{Results and Discussion}
\subsection{Experiment Metrics} 
\begin{figure}[!ht]
    \centering
    \begin{subfigure}[t]{0.33\textwidth}
        \includegraphics[width=\textwidth]{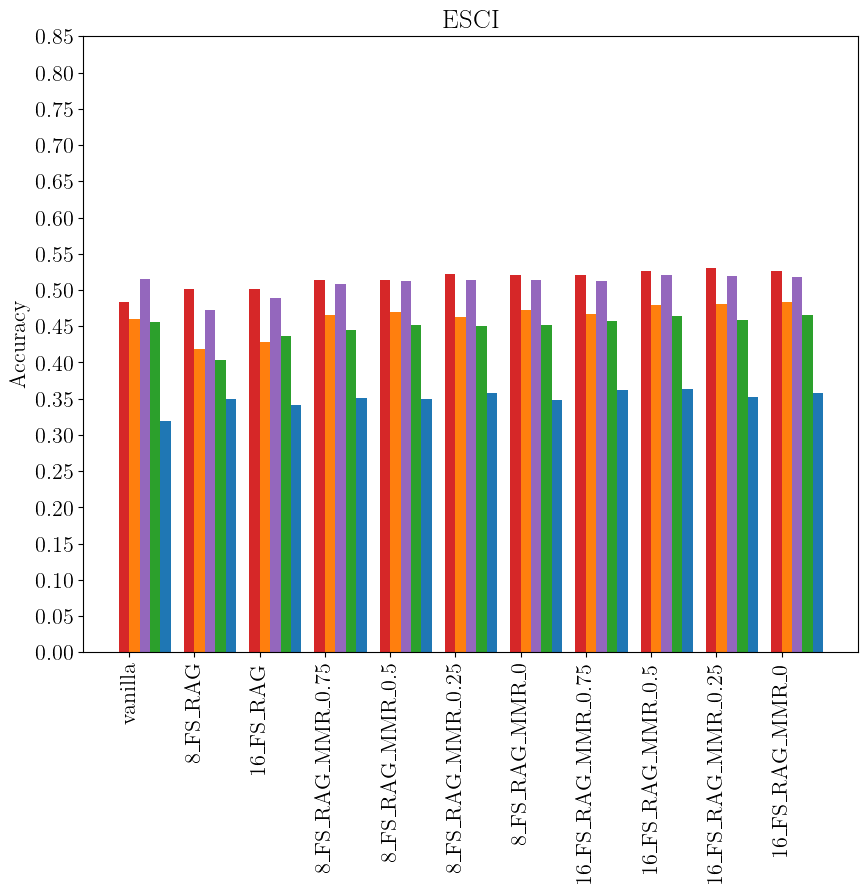}
        \caption{Accuracy of experiments on ESCI data}
    \end{subfigure}

    \begin{subfigure}[t]{0.33\textwidth}
        \includegraphics[width=\textwidth]{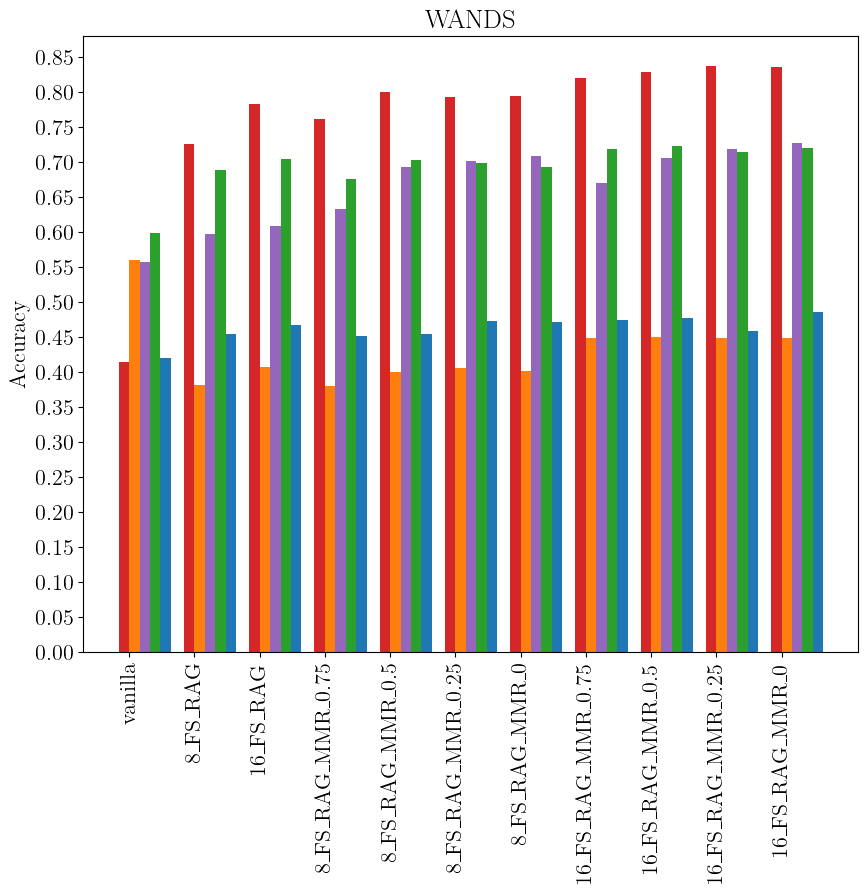}
        \caption{Accuracy of experiments on WANDS data}
    \end{subfigure}
    
    \begin{subfigure}[t]{0.33\textwidth}
        \includegraphics[width=\textwidth]{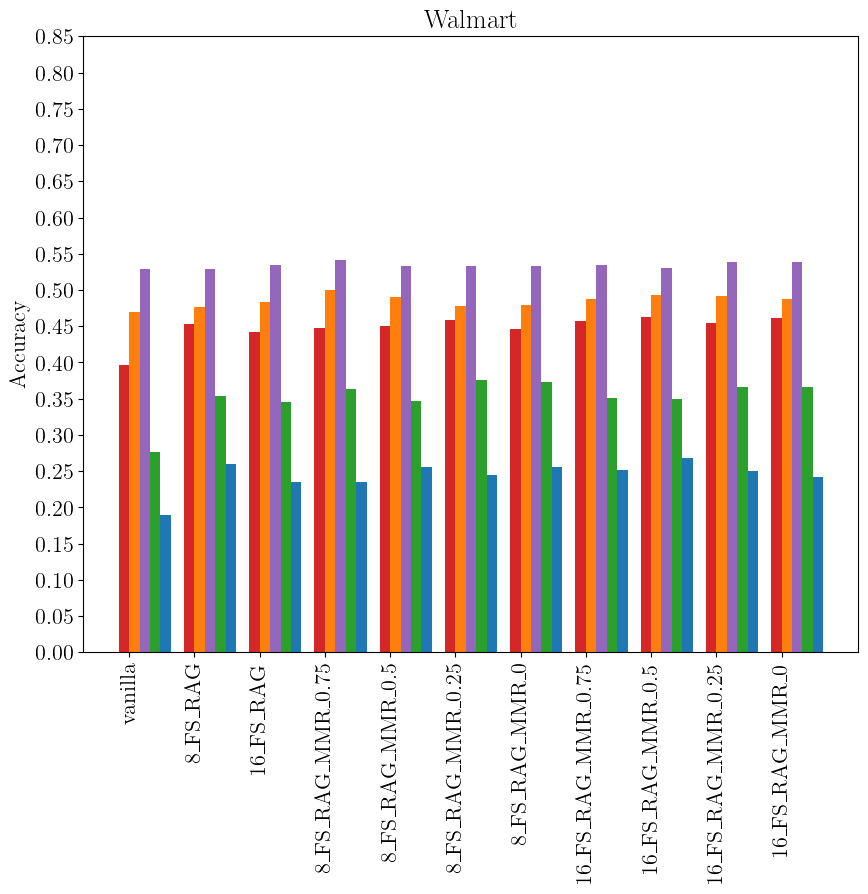}
        \caption{Accuracy of experiments on Walmart data}
    \end{subfigure}
    \caption{Color Code: \usellmone{$\blacksquare$ LLM1}, \usellmtwo{$\blacksquare$ LLM2}, \usellmthree{$\blacksquare$ LLM3},\\ \usellmfour{$\blacksquare$ LLM4},\usellmfive{$\blacksquare$ LLM5}}
    \label{figure:graphs}
\end{figure}

Across experiments on all three datasets, we observe LLM2 and LLM5 outperform all other models (Please refer to Fig \ref{figure:graphs}, and for the numeric outputs in Table~\ref{tab:results}). Our results show that providing FS examples improves accuracy over zero-shot in most cases, and RAG FS offers further improvements in accuracy metrics in some instances (see Table~\ref{tab:results}). However, we observe that the best accuracy is obtained when RAG is used in combination with diversification by MMR, which we believe is due to how the contextual information exploited by the different LLMs.
Further, for experiments with 8 and 16 few shot examples using RAG to select examples, we tried various values of $\lambda_{MMR} \in [0.75,0.5,0.25,0.0]$ . Setting $\lambda_{MMR} = 1$ is the same as the FS\_RAG experiments because retrieval is solely based on relevance thus inducing zero diversity.
We observe that for different datasets, and different values of FS, the optimal value of $\lambda_{MMR}$ up to the resolution of our grid (0.25), can be different. Overall, we observe that diversity helps in improving accuracy. We observe that increasing diversification i.e. lower values of $\lambda_{MMR}$ improves accuracy metrics, as shown in Fig \ref{figure:MMRgraphs}. This is because the retrieved examples may have some degree of semantic overlap, hence diversification helps add novel information within the prompt. 
We also include sample prompts from our experiments with WANDS data in the Appendix A1 and A2.
A sample set of Q-P pairs on the ESCI data is shown in Table \ref{table:examplesESCI} and a sample set of Q-P pairs on the WANDS data is shown in Table \ref{table:examplesWANDS}. For some predictions that do not match the ground truth, such as Example \#2 in Table \ref{table:examplesESCI} and Example \#4 in Table \ref{table:examplesWANDS}, the prediction labels are arguably more accurate than the ground truth labels. We can potentially use this approach to improve these datasets by identifying errors in the existing human-annotated ground truth labels.
For Walmart Mexico data, LLM5 with 16 FS\_RAG\_MMR with $\lambda_{MMR}=0.75$ approaches human-level weighted F1 Score (see Table~\ref{tab:results}).
\begin{table}[!h]
    \fontsize{7}{9}\selectfont
  \caption{Example results on ESCI data}
\begin{tabular}{cp{4cm}cc}
\toprule
\# & \textbf{Query-Product pair} & \textbf{Ground Truth} & \textbf{Prediction} \\ \hline
1 & \textbf{Q:} \texttt{womens knee high socks} \newline
\textbf{P:} \texttt{12 Pairs Women Knee High Socks 196013} & Exact & Exact \\ \hline 
2 & \textbf{Q:} \texttt{bar style dartboard} \newline
\textbf{P:}\hspace{0.05in}\texttt{Viper EVA V-Foam Dartboard Surround Wall Protector, Black} & Exact & Complement \\ \hline
3 & \textbf{Q:} \texttt{fiskars axe sharpener} \newline
\textbf{P:} \texttt{Fiskars X25 Splitting Axe, $28$-Inch} & Complement & Irrelevant \\ \hline
4 & \textbf{Q:} \texttt{helium tank}  \newline
\textbf{P:} \texttt{Helium Tank with 50 Balloons and White Ribbon} & Substitute & Exact \\
\bottomrule
\end{tabular}
\label{table:examplesESCI}
\end{table}
\begin{table}[!h]
    \fontsize{7}{9}\selectfont
  \caption{Example results on WANDS data}
\begin{tabular}{cp{4cm}cc}
\toprule
\textbf{\#} & \textbf{Query-Product pair} & \textbf{Ground Truth} & \textbf{Prediction} \\ \hline 
1 & \textbf{Q:} \texttt{leather chair} \newline
\textbf{P:}\hspace{0.05in}\texttt{31" wide top grain leather armchair} & Exact & Exact \\ \hline
2 & \textbf{Q:} \texttt{coffee container} \newline
\textbf{P:}\hspace{0.05in}\texttt{gillis coffee table with storage} & Irrelevant & Irrelevant \\ \hline
3 & \textbf{Q:} \texttt{basket planter} \newline
\textbf{P:}\hspace{0.05in}\texttt{classic rolled rim stone pot planter} & Partial & Partial \\ \hline
4 & \textbf{Q:} \texttt{wine bar} \newline
\textbf{P:}\hspace{0.05in}\texttt{corene sixteenth century italian replica mini bar} & Partial & Exact \\ \hline
5 & \textbf{Q:} \texttt{unique coffee tables} \newline
\textbf{P:} \texttt{solid wood block coffee table} & Partial & Exact \\ \hline
6 & \textbf{Q:} \texttt{beaded curtains} \newline
\textbf{P:} \texttt{fade single curtain rod} & Irrelevant & Partial \\
\bottomrule
\end{tabular}
\label{table:examplesWANDS}
\end{table}

\subsection{Time and Cost Analysis}
External human-annotations had a turnaround time of about 2-3 weeks. Labelers often take about a week to familiarize themselves with the task requirements, and then another 2 weeks to manually label the 1000 records (Q-P pairs) they are provided. Therefore, turnaround for third party human-annotations is estimated to be about 30 mins/record. In contrast, our experiments show that LLMs do the task in a fraction of the time. Our best performing configuration for LLM5  takes a little over 3 hours to complete, so the upper bound on LLMs for this annotation task is about 8.3 seconds/record. However, with open-source models most experiments ran in under 5 minutes on a GPU machine with an inference time of about 0.3 seconds/record. These models demonstrate a significantly efficient run time for large datasets such as those from Walmart's e-commerce ecosystem, thereby offering a viable alternative. Moreover, the open-source nature of these models confers an added advantage in terms of flexibility, allowing for more customized adjustments according to specific requirements through fine-tuning.

The costs of using LLMs are also significantly less than third-party human labelers. According to our estimates, labeling each Q-P pair is about 500 times cheaper with LLMs than human labelers, even when considering compute resources and API costs. LLMs have the potential to be even more cost-effective with techniques like batching of inputs/outputs, and conciseness of prompts. Open-source models may be a more suitable choice over proprietary LLMs in many situations since they can be fine-tuned on custom data to achieve better performance and mitigate any potential data privacy/leakage concerns.

\section{Conclusion}
Our work demonstrates the potential of LLMs in providing accurate and reliable query-product relevance data at scale. By leveraging LLMs with the state-of-the-art prompt engineering techniques, we are able to achieve accuracy comparable to human-labelled query-product pairs, while significantly reducing the time and cost required. Our experiments demonstrated the novel use of MMR in RAG for query-product relevance labeling has shown good improvements and strong promise for future work. We also showed how  techniques like Chain-of-Thought and Few Shot prompting can help improve the accuracy of LLMs for this task. This scalable alternative to traditional human-annotation methods has far-reaching implications for information retrieval domains such as search and recommendation. One could consider using product descriptions and other attributes that are often available in the product catalogs and taxonomies, however, they are often noisy and would increase the number of input tokens to the LLMs which would have additional computational cost. Our productionized approach within Walmart Global Tech has already yielded a large number of query-product relevance labels, which are currently being used to evaluate improvements to various search algorithms. This not only improves the efficiency of our search systems but also enables us to provide better user experiences through more relevant search results. Furthermore, our method's broad applicability across various information retrieval domains makes it a valuable tool for optimizing products and content for greater user interactions and revenue generation.



\bibliographystyle{ACM-Reference-Format}
\bibliography{references}


\appendix
\section{Appendix}
\subsection{Sample prompt : WANDS Dataset - 8 FS with RAG}
\begin{minipage}{3.4in}
{\tiny{
\begin{lstlisting}
You are a search engine in an eCommerce website. For a given customer query and a product title, please annotate each product title in the list as one of these options: 'Exact', 'Partial', 'Irrelevant'.
Exact : this label represents the surfaced product fully matches the search query.
Partial : this label represents the surfaced product that does not fully match the search query. It only matches the target entity of the query, but does not satisfy the modifiers for the query.
Irrelevant : this label indicates the product is not relevant to the query.
The response should be in a python dictionary format {"rating":label}, where label which is either 'Exact', 'Partial', or 'Irrelevant'.

#### Here are some examples:

query: wood coffee table set by storage, product title: coffee table
{'rating': 'Partial'}

query: wood coffee table set by storage, product title: coffee table with storage
{'rating': 'Partial'}

query: wood coffee table set by storage, product title: onshuntay coffee table
{'rating': 'Partial'}

query: wood coffee table set by storage, product title: wooden coffee table
{'rating': 'Partial'}

query: wood coffee table set by storage, product title: wood coffee table
{'rating': 'Partial'}

query: wood coffee table set by storage, product title: ahern coffee table
{'rating': 'Partial'}

query: wood coffee table set by storage, product title: fromm wood table
{'rating': 'Partial'}

query: wood coffee table set by storage, product title: bahareh coffee table
{'rating': 'Partial'}

Now rate the relevance of this pair:
query: wood coffee table set by storage, product title: mikell 2 piece coffee table set
\end{lstlisting}
}}
\end{minipage}
In the prompt above, there are some examples from RAG that have a high degree of overlap like the product titles "wooden coffee table" and "wood coffee table". This issue is overcome by MMR-based diversity, as seen in the next prompt

\subsection{Sample prompt : WANDS Dataset - 8 FS RAG MMR $\lambda_{MMR} = 0$}
\begin{minipage}{3.4in}
{\tiny{
\begin{lstlisting}
You are a search engine in an eCommerce website. For a given customer query and a product title, please annotate each product title in the list as one of these  options: 'Exact', 'Partial', 'Irrelevant'
Exact : this label represents the surfaced product fully 
matches the search query.
Partial : this label represents the surfaced product that
does not fully match the search query. It only
matches the target entity of the query, but does not 
satisfy the modifiers for the query.
Irrelevant : this label indicates the product is not relevant to the query.
The response should be in a python dictionary format {"rating":label},
where label which is either 'Exact', 'Partial', or 'Irrelevant'.

#### Here are some examples:

query: wood coffee table set by storage, product title: coffee table
{'rating': 'Partial'}

query: wood coffee table set by storage, product title: gabrielle end table storage
{'rating': 'Partial'}

query: wood coffee table set by storage, product title: mylor solid wood coffee table
{'rating': 'Partial'}

query: wood coffee table set by storage, product title: oday solid wood coffee table with storage
{'rating': 'Partial'}

query: wood coffee table set by storage, product title: radford coffee table with storage
{'rating': 'Partial'}

query: wood coffee table set by storage, product title: berg solid coffee table with storage
{'rating': 'Partial'}

query: wood coffee table set by storage, product title: aule solid wood end table with storage
{'rating': 'Partial'}

query: wood coffee table set by storage, product title: hedda coffee table
{'rating': 'Partial'}

Now rate the relevance of this pair:
query: wood coffee table set by storage, product title: mikell 2 piece coffee table set
\end{lstlisting}
}
}
\end{minipage}

\subsection{Experimental results}
\tiny
\begin{table}[H]
\begin{tabular}{l|ccc|ccc|ccc}
Model & \multicolumn{3}{c}{ESCI} & \multicolumn{3}{c}{WANDS} & \multicolumn{3}{c}{Walmart}\\ 
\\ &  Acc. & $F_{1}$ & $F^{\omega}_{1}$ & Acc. & $F_{1}$ & $F^{\omega}_{1}$ & Acc. & $F_{1}$ & $F^{\omega}_{1}$ \\
\midrule
3rd-party labelers & - & - & - & - & - & - & 0.648 & 0.615 & 0.452\\\hline
LLM1 + VANILLA & 0.460 & 0.340 & 0.434 & 0.560 & 0.522 & 0.567 & 0.469 & 0.208 & 0.442\\
LLM1 + 8\_FS & 0.463 & 0.335 & 0.430 & 0.468 & 0.463 & 0.480 & 0.494 & 0.204 & 0.447\\
LLM1 + 16\_FS & 0.466 & 0.343 & 0.427 & 0.443 & 0.444 & 0.449 & 0.479 & 0.196 & 0.442\\
LLM1 + 8\_FS\_RAG & 0.419 & 0.342 & 0.423 & 0.381 & 0.376 & 0.353 & 0.476 & 0.208 & 0.445 \\
LLM1 + 16\_FS\_RAG & 0.428 & 0.348 & 0.423 & 0.407 & 0.406 & 0.380 & 0.484 & 0.199 & 0.442 \\
LLM1 + 8\_FS\_COT & 0.470 & 0.346 & 0.444 & 0.461 & 0.455 & 0.476 & 0.471 & 0.175 & 0.420 \\
LLM1 + 16\_FS\_COT & 0.460 & 0.330 & 0.424 & 0.439 & 0.441 & 0.450 & 0.487 & 0.197 & 0.439 \\
LLM1 + 8\_FS\_RAG\_COT & 0.428 & 0.348 & 0.433 & 0.378 & 0.368 & 0.356 & 0.485 & 0.217 & 0.453 \\
LLM1 + 16\_FS\_RAG\_COT & 0.429 & 0.345 & 0.426 & 0.436 & 0.430 & 0.420 & 0.457 & 0.190 & 0.426 \\
LLM1 + 8\_FS\_RAG\_MMR\_0.75 & 0.466 & 0.363 & 0.455 & 0.381 & 0.379 & 0.363 & 0.500 & 0.297 & 0.458 \\
LLM1 + 8\_FS\_RAG\_MMR\_0.5 & 0.469 & 0.363 & 0.448 & 0.400 & 0.398 & 0.398 & 0.490 & 0.278 & 0.447 \\
LLM1 + 8\_FS\_RAG\_MMR\_0.25 & 0.463 & 0.351 & 0.443 & 0.406 & 0.408 & 0.403 & 0.478 & 0.268 & 0.437 \\
LLM1 + 8\_FS\_RAG\_MMR\_0 & 0.473 & 0.358 & 0.450 & 0.402 & 0.402 & 0.399 & 0.479 & 0.280 & 0.443 \\
LLM1 + 16\_FS\_RAG\_MMR\_0.75 & 0.467 & 0.374 & 0.453 & 0.449 & 0.449 & 0.438 & 0.488 & 0.274 & 0.454 \\
LLM1 + 16\_FS\_RAG\_MMR\_0.5 & 0.479 & 0.365 & 0.459 & 0.450 & 0.451 & 0.445 & 0.493 & 0.291 & 0.454 \\
LLM1 + 16\_FS\_RAG\_MMR\_0.25 & 0.481 & 0.371 & 0.456 & 0.449 & 0.452 & 0.439 & 0.492 & 0.260 & 0.446 \\
LLM1 + 16\_FS\_RAG\_MMR\_0 & 0.484 & 0.368 & 0.457 & 0.449 & 0.452 & 0.438 & 0.488 & 0.283 & \textbf{0.455}  \\
\hline
LLM2 + VANILLA & 0.515 & 0.367 & 0.469 & 0.557 & 0.531 & 0.568 & 0.529 & 0.159 & 0.409\\
LLM2 + 8\_FS & 0.522 & 0.410 & 0.499 & 0.542 & 0.515 & 0.555 & 0.508 & 0.175 & 0.427\\
LLM2 + 16\_FS & 0.520 & 0.395 & 0.494 & 0.521 & 0.504 & 0.538 & 0.507 & 0.174 & 0.424\\
LLM2 + 8\_FS\_RAG & 0.472 & 0.388 & 0.464 & 0.597 & 0.571 & 0.611 & 0.529 & 0.164 & 0.415\\
LLM2 + 16\_FS\_RAG & 0.489 & 0.387 & 0.477 & 0.609 & 0.582 & 0.627 & 0.535 & 0.170 & 0.419\\
LLM2 + 8\_FS\_COT & 0.522 & 0.401 & 0.504 & 0.508 & 0.496 & 0.519 & 0.509 & 0.170 & 0.423\\
LLM2 + 16\_FS\_COT & 0.517 & 0.403 & 0.497 & 0.511 & 0.501 & 0.524 & 0.501 & 0.174 & 0.418\\
LLM2 + 8\_FS\_RAG\_COT & 0.460 & 0.366 & 0.452 & 0.599 & 0.572 & 0.613 & 0.526 & 0.151 & 0.403\\
LLM2 + 16\_FS\_RAG\_COT & 0.476 & 0.387 & 0.464 & 0.625 & 0.598 & 0.639 & 0.533 & 0.164 & 0.413\\
LLM2 + 8\_FS\_RAG\_MMR\_0.75 & 0.509 & 0.408 & 0.498 & 0.633 & 0.604 & 0.648 & 0.541 & 0.227 & 0.422\\
LLM2 + 8\_FS\_RAG\_MMR\_0.5 & 0.512 & 0.409 & 0.499 & 0.692 & 0.654 & 0.704 & 0.533 & 0.215 & 0.415\\
LLM2 + 8\_FS\_RAG\_MMR\_0.25 & 0.513 & 0.399 & 0.499 & 0.701 & 0.660 & 0.711 & 0.533 & 0.215 & 0.409\\
LLM2 + 8\_FS\_RAG\_MMR\_0 & 0.514 & 0.408 & 0.499 & 0.708 & 0.666 & 0.718 & 0.534 & 0.212 & 0.414\\
LLM2 + 16\_FS\_RAG\_MMR\_0.75 & 0.512 & 0.411 & 0.498 & 0.669 & 0.639 & 0.686 & 0.535 & 0.227 & 0.415\\
LLM2 + 16\_FS\_RAG\_MMR\_0.5 & 0.521 & 0.410 & 0.507 & 0.706 & 0.664 & 0.719 & 0.531 & 0.213 & 0.408\\
LLM2 + 16\_FS\_RAG\_MMR\_0.25 & 0.519 & 0.407 & 0.502 & 0.718 & 0.678 & 0.730 & 0.539 & 0.215 & 0.422\\
LLM2 + 16\_FS\_RAG\_MMR\_0 & 0.518 & 0.406 & 0.502 & 0.726 & 0.687 & 0.738 & 0.539 & 0.225 & 0.419\\
\hline
LLM3 + VANILLA & 0.456 & 0.394 & 0.469 & 0.599 & 0.535 & 0.601 & 0.276 & 0.171 & 0.259\\
LLM3 + 8\_FS & 0.426 & 0.374 & 0.445 & 0.637 & 0.561 & 0.636 & 0.225 & 0.140 & 0.195\\
LLM3 + 16\_FS & 0.436 & 0.377 & 0.456 & 0.658 & 0.567 & 0.652 & 0.224 & 0.126 & 0.208\\
LLM3 + 8\_FS\_RAG & 0.404 & 0.360 & 0.411 & 0.688 & 0.571 & 0.673 & 0.354 & 0.200 & 0.361\\
LLM3 + 16\_FS\_RAG & 0.437 & 0.386 & 0.437 & 0.705 & 0.573 & 0.683 & 0.345 & 0.193 & 0.353\\
LLM3 + 8\_FS\_COT & 0.428 & 0.372 & 0.447 & 0.651 & 0.541 & 0.636 & 0.233 & 0.132 & 0.186\\
LLM3 + 16\_FS\_COT & 0.433 & 0.371 & 0.450 & 0.664 & 0.530 & 0.640 & 0.220 & 0.121 & 0.182\\
LLM3 + 8\_FS\_RAG\_COT & 0.408 & 0.356 & 0.406 & 0.680 & 0.511 & 0.639 & 0.379 & 0.190 & 0.384\\
LLM3 + 16\_FS\_RAG\_COT & 0.424 & 0.358 & 0.411 & 0.692 & 0.512 & 0.650 & 0.379 & 0.191 & 0.387\\
LLM3 + 8\_FS\_RAG\_MMR\_0.75 & 0.445 & 0.390 & 0.448 & 0.675 & 0.543 & 0.654 & 0.364 & 0.241 & 0.371\\
LLM3 + 8\_FS\_RAG\_MMR\_0.5 & 0.452 & 0.393 & 0.456 & 0.703 & 0.580 & 0.679 & 0.347 & 0.234 & 0.353\\
LLM3 + 8\_FS\_RAG\_MMR\_0.25 & 0.451 & 0.394 & 0.455 & 0.698 & 0.569 & 0.674 & 0.376 & 0.260 & 0.381\\
LLM3 + 8\_FS\_RAG\_MMR\_0 & 0.452 & 0.394 & 0.457 & 0.693 & 0.565 & 0.670 & 0.373 & 0.255 & 0.382\\
LLM3 + 16\_FS\_RAG\_MMR\_0.75 & 0.457 & 0.398 & 0.451 & 0.719 & 0.580 & 0.693 & 0.351 & 0.222 & 0.354\\
LLM3 + 16\_FS\_RAG\_MMR\_0.5 & 0.464 & 0.400 & 0.461 & 0.722 & 0.598 & 0.697 & 0.350 & 0.264 & 0.359\\
LLM3 + 16\_FS\_RAG\_MMR\_0.25 & 0.459 & 0.407 & 0.457 & 0.714 & 0.585 & 0.688 & 0.367 & 0.257 & 0.376\\
LLM3 + 16\_FS\_RAG\_MMR\_0 & 0.465 & 0.406 & 0.462 & 0.720 & 0.591 & 0.695 & 0.367 & 0.255 & 0.373\\
\hline
LLM4 + VANILLA & 0.319 & 0.259 & 0.329 & 0.420 & 0.367 & 0.443 & 0.189 & 0.123 & 0.245 \\
LLM4 + 8\_FS & 0.318 & 0.256 & 0.333 & 0.392 & 0.362 & 0.420 & 0.229 & 0.129 & 0.273 \\
LLM4 + 16\_FS & 0.339 & 0.272 & 0.352 & 0.396 & 0.371 & 0.425 & 0.196 & 0.117 & 0.244 \\
LLM4 + 8\_FS\_RAG & 0.349 & 0.290 & 0.361 & 0.454 & 0.402 & 0.483 & 0.259 & 0.172 & 0.314 \\
LLM4 + 16\_FS\_RAG & 0.342 & 0.277 & 0.351 & 0.468 & 0.420 & 0.495 & 0.236 & 0.145 & 0.288 \\
LLM4 + 8\_FS\_COT & 0.321 & 0.264 & 0.339 & 0.356 & 0.332 & 0.388 & 0.227 & 0.122 & 0.268 \\
LLM4 + 16\_FS\_COT & 0.315 & 0.253 & 0.329 & 0.369 & 0.339 & 0.402 & 0.192 & 0.117 & 0.236 \\
LLM4 + 8\_FS\_RAG\_COT & 0.331 & 0.270 & 0.342 & 0.437 & 0.388 & 0.467 & 0.250 & 0.152 & 0.301 \\
LLM4 + 16\_FS\_RAG\_COT & 0.353 & 0.283 & 0.362 & 0.461 & 0.410 & 0.491 & 0.244 & 0.140 & 0.289 \\
LLM4 + 8\_FS\_RAG\_MMR\_0.75 & 0.350 & 0.282 & 0.362 & 0.452 & 0.402 & 0.479 & 0.235 & 0.146 & 0.286 \\
LLM4 + 8\_FS\_RAG\_MMR\_0.5 & 0.349 & 0.275 & 0.359 & 0.454 & 0.402 & 0.482 & 0.255 & 0.160 & 0.307 \\
LLM4 + 8\_FS\_RAG\_MMR\_0.25 & 0.358 & 0.288 & 0.368 & 0.473 & 0.421 & 0.500 & 0.245 & 0.146 & 0.302 \\
LLM4 + 8\_FS\_RAG\_MMR\_0 & 0.348 & 0.289 & 0.355 & 0.472 & 0.420 & 0.498 & 0.256 & 0.150 & 0.315 \\
LLM4 + 16\_FS\_RAG\_MMR\_0.75 & 0.362 & 0.292 & 0.369 & 0.475 & 0.432 & 0.501 & 0.251 & 0.149 & 0.303 \\
LLM4 + 16\_FS\_RAG\_MMR\_0.5 & 0.363 & 0.282 & 0.369 & 0.477 & 0.435 & 0.505 & 0.268 & 0.159 & 0.323 \\
LLM4 + 16\_FS\_RAG\_MMR\_0.25 & 0.353 & 0.280 & 0.361 & 0.459 & 0.414 & 0.489 & 0.250 & 0.158 & 0.303 \\
LLM4 + 16\_FS\_RAG\_MMR\_0 & 0.358 & 0.287 & 0.366 & 0.485 & 0.439 & 0.511 & 0.243 & 0.151 & 0.293 \\
\hline
LLM5 + VANILLA & 0.483 & 0.417 & 0.489 & 0.415 & 0.368 & 0.383 & 0.397 & 0.231 & 0.437\\
LLM5 + 8\_FS & 0.514 & 0.454 & 0.519 & 0.635 & 0.612 & 0.643 & 0.359 & 0.203 & 0.394\\
LLM5 + 16\_FS & 0.516 & 0.447 & 0.524 & 0.656 & 0.633 & 0.663 & 0.344 & 0.199 & 0.386\\
LLM5 + 8\_FS\_RAG & 0.501 & 0.442 & 0.507 & 0.726 & 0.683 & 0.728 & 0.453 & 0.236 & 0.481\\
LLM5 + 16\_FS\_RAG & 0.501 & 0.446 & 0.507 & 0.783 & 0.741 & 0.783 & 0.443 & 0.237 & 0.474\\
LLM5 + 8\_FS\_COT & 0.490 & 0.431 & 0.499 & 0.647 & 0.617 & 0.653 & 0.362 & 0.207 & 0.395\\
LLM5 + 16\_FS\_COT & 0.483 & 0.439 & 0.492 & 0.663 & 0.637 & 0.668 & 0.322 & 0.176 & 0.360\\
LLM5 + 8\_FS\_RAG\_COT & 0.477 & 0.435 & 0.483 & 0.730 & 0.691 & 0.732 & 0.438 & 0.235 & 0.466\\
LLM5 + 16\_FS\_RAG\_COT & 0.458 & 0.428 & 0.463 & 0.715 & 0.683 & 0.719 & 0.399 & 0.227 & 0.438\\
LLM5 + 8\_FS\_RAG\_MMR\_0.75 & 0.514 & 0.457 & 0.518 & 0.761 & 0.720 & 0.762 & 0.448 & 0.240 & 0.475\\
LLM5 + 8\_FS\_RAG\_MMR\_0.5 & 0.514 & 0.464 & 0.518 & 0.799 & 0.761 & 0.799 & 0.451 & 0.244 & 0.481\\
LLM5 + 8\_FS\_RAG\_MMR\_0.25 & 0.523 & 0.453 & 0.526 & 0.792 & 0.753 & 0.792 & 0.458 & 0.247 & 0.488\\
LLM5 + 8\_FS\_RAG\_MMR\_0 & 0.520 & 0.454 & 0.524 & 0.795 & 0.750 & 0.794 & 0.446 & 0.242 & 0.476\\
LLM5 + 16\_FS\_RAG\_MMR\_0.75 & 0.521 & 0.464 & 0.526 & 0.820 & 0.786 & 0.821 & 0.457 & 0.244 & 0.487\\
LLM5 + 16\_FS\_RAG\_MMR\_0.5 & 0.526 & 0.463 & 0.530 & 0.828 & 0.790 & 0.828 & 0.462 & 0.252 & 0.492\\
LLM5 + 16\_FS\_RAG\_MMR\_0.25 & 0.530 & 0.467 & 0.533 & 0.837 & 0.800 & 0.837 & 0.454 & 0.243 & 0.483\\
LLM5 + 16\_FS\_RAG\_MMR\_0 & 0.526 & 0.456 & 0.529 & 0.835 & 0.796 & 0.834 & 0.461 & 0.247 & \textbf{0.489}\\
\bottomrule
\end{tabular}
\caption{Experimental results}
\label{tab:results}
\end{table}

\subsection{Plots for MMR Experiments}
\begin{figure}[H]
    \centering
    \begin{subfigure}[t]{0.33\textwidth}
        \includegraphics[width=0.8\textwidth]{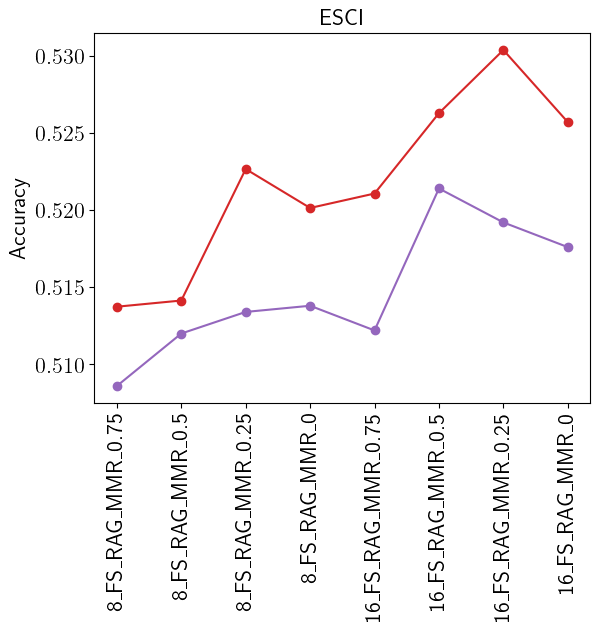}
        \caption{ESCI MMR experiments}
    \end{subfigure}
    \begin{subfigure}[t]{0.33\textwidth}
        \includegraphics[width=0.8\textwidth]{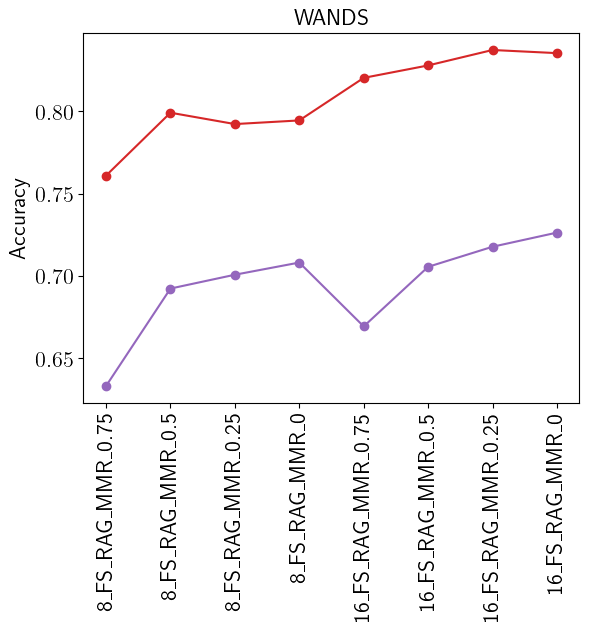}[
        \caption{WANDS MMR experiments}
    \end{subfigure}
    \begin{subfigure}[t]{0.33\textwidth}
        \includegraphics[width=0.8\textwidth]{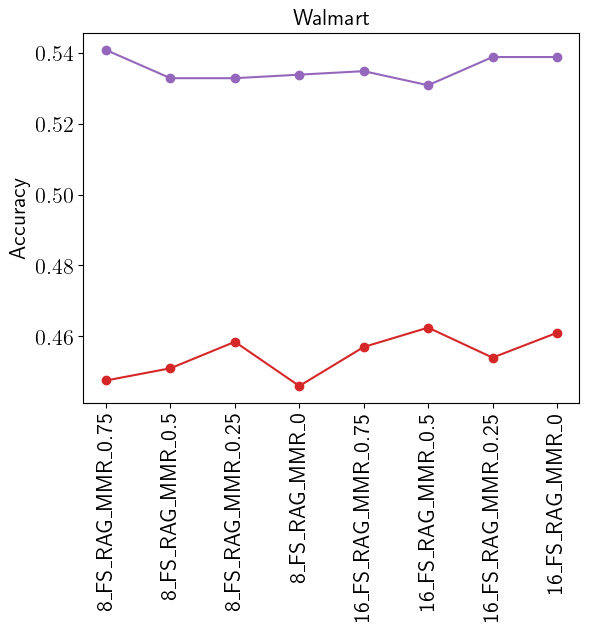}
        \caption{Walmart MMR experiments}
    \end{subfigure}
    \caption{Color Code: \usellmtwo{$\blacksquare$ LLM2}, \usellmfive{$\blacksquare$ LLM5}}
    \label{figure:MMRgraphs}
\end{figure}

\end{document}